\documentclass[useAMS,usenatbib]{mn2e}

\usepackage{amssymb}
\usepackage{amsmath}

\newcommand{\HIbold}{{H\footnotesize{\bf{I}}}}
\newcommand{\HI}{\mbox{\sc H{i}}}
\newcommand{\HII}{\mbox{\sc H{ii}}}
\newcommand{\nhi}{\mbox{$N_{\rm HI}$}}
\newcommand{\Lya}{\mbox{Ly$\alpha$}}
\newcommand{\msun}{\mbox{$M_\odot$}}
\newcommand{\mhi}{\mbox{$M_{\rm HI}$}}
\newcommand{\kms}{\mbox{km s$^{-1}$}}
\newcommand{\cm}{${\rmn{cm}}^{-2}$}
\newcommand{\lsun}{\mbox{$L_\odot$}}
\newcommand{\mhilb}{\mbox{$M_{\rm HI}$-to-$L_B$}}

\newcommand{\surf}{$\overline{\mu}_{25}$}
\newcommand{\dndz}{$dN_{DLA}/dz$}
\newcommand{\rhohi}{\mbox{$\rho_{\rm HI}$}}

\title{The Column Density Distribution Function at \boldmath{$z=0$}
 from \HIbold\ Selected Galaxies.}
\author[E. Ryan-Weber]
       {Emma V. Ryan-Weber,$^{1,2}$ Rachel L. Webster$^1$ and Lister 
	Staveley-Smith$^2$\\
        $^1$School of Physics, Univeristy of Melbourne, VIC 3010 Australia\\
	$^2$Australia Telescope National Facility, CSIRO, PO Box 76, 
	Epping, NSW 1710, Australia}

\begin{document}

\date{Accepted ***. Received ***; in original form ***}

\pagerange{\pageref{firstpage}--\pageref{lastpage}} \pubyear{2003}

\maketitle
 
\label{firstpage}

\begin{abstract}
We have measured the column density distribution function, f(\nhi), at
$z=0$ using 21-cm \HI\ emission from galaxies selected from a blind
\HI\ survey. f(\nhi) is found to be smaller and flatter at $z=0$ than
indicated by high-redshift measurements of Damped Lyman-$\alpha$ (DLA)
systems, consistent with the predictions of hierarchical galaxy
formation. The derived DLA number density per unit redshift, \dndz\
=$0.058$, is in moderate agreement with values calculated from
low-redshift QSO absorption line studies. We use two different methods
to determine the types of galaxies which contribute most to the DLA
cross-section: comparing the power law slope of f(\nhi) to theoretical
predictions and analysing contributions to \dndz. We find that
comparison of the power law slope cannot rule out spiral discs as the
dominant galaxy type responsible for DLA systems. Analysis of \dndz\
however, is much more discriminating. We find that galaxies with $\log
\mhi< 9.0$ make up 34\% of \dndz; Irregular and Magellanic types
contribute 25\%; galaxies with surface brightness, \surf $>$ 24 mag
arcsec$^{-2}$ account for 22\% and sub-$L_*$ galaxies contribute 45\%
to \dndz. We conclude that a large range of galaxy types give rise to
DLA systems, not just large spiral galaxies as previously speculated.
\end{abstract}


\section{Introduction}
Lyman-$\alpha$ (\Lya) absorption systems, in the continuum spectra of
background QSOs, provide a powerful probe of the distribution of
neutral hydrogen in the Universe. Damped \Lya\ (DLA) systems have
column densities \nhi~$\geq 2\times10^{20}$\cm\ and contain the bulk
of the neutral gas in the Universe \citep{Storrie00}. The value of
$\Omega_{gas}$ from DLAs at high redshift agrees well with
$\Omega_{star}$ in present day galaxies. The fact that $\Omega_{gas}$
decreases monotonically with decreasing redshift and approaches
$\Omega_{gas}(z=0)$ from 21-cm emission of local galaxies also
provides strong evidence that DLAs are caused by galaxies
\citep{Wolfe95}. The properties of these galaxies however, remain
mostly unconstrained.

There are two primary competing theories on the evolution of neutral
gas in galaxies. The first is monolithic collapse of protogalactic
discs \citep{Wolfe86}. This scenario, supported by \cite{Prochaska97},
proposes that the kinematics of neutral gas tracing-ions in DLAs are
best described by thick, rotating galactic discs. Alternatively,
neutral gas in galaxies could be accreted hierarchically from
sub-galactic halos. \cite{Haehnelt98} argue that the kinematics of
DLAs can equally be explained by sight-lines through irregular
protogalactic clumps.

Since massive spiral galaxies contain the bulk of the \HI\ at $z=0$,
it has been presumed that they also account for most of the
\nhi~$\geq2\times10^{20}$\cm\ cross-section \citep[e.g.][]{Wolfe86,
Rao93}. With the advent of space-based UV spectroscopy, and the
discovery of low redshift ($z<1.65$) DLAs, this assumption can now be
tested. At low redshift the galaxies responsible for DLAs can be
resolved, so their luminosities, surface brightness and
morphological types can be determined.

At $z<1.65$ fewer than 30 DLAs are known. Images of 15 DLA galaxies
have been published. The morphological breakdown includes 7 low
surface brightness (LSB) galaxies, 6 spirals and 2 compact dwarf-like
galaxies
\citep{LeBrun97,pettini00,Turnshek01,Bowen01a,Nestor02}. These
galaxies are distributed evenly in luminosity, with 7 galaxies having
$L<L_*$. Only five of these galaxies have measured spectroscopic
redshifts. Unless the galaxy has a confirmed redshift matching that of
the absorption systems, it is not certain that a particular galaxy is
responsible for a DLA system.

To determine the \HI\ emission properties of these DLA galaxies, they
need to have even lower redshifts (due to the weakness of the 21-cm
transition). There is only one case of \HI\ emission detected from a
DLA galaxy. The DLA galaxy SBS 1543+593 \citep{Bowen01a} has been
detected \citep{Bowen01b} and mapped \citep{Chengalur02} in 21-cm. The
candidate (\Lya\ column density yet to be published) DLA galaxy NGC
4203 reported by \cite{Miller99} has also been mapped in 21-cm
\citep{vandriel88}.

For a set of random QSO sight-lines, DLA absorbers simply map out the
cross-section on the sky with \nhi~$\geq2\times10^{20}$\cm. For this
reason, it is not surprising to find a variety of galaxies responsible
for DLAs. There are also theoretical suggestions that different galaxy
types, such as LSB galaxies may contribute significantly to the high
\nhi\ cross-section \citep{Jimenez99, Boissier02}. One way to
characterize this cross-section is to use the column density
distribution function, $f(\nhi)$.

$f(\nhi)$ is defined as the number of absorbers in the range \nhi\ to
$\nhi + d\nhi$ per unit \nhi\ per unit absorption distance. The most
recent determination of $f(\nhi)$ over a range of redshifts, using a
statistical sample of 646 QSOs, with 85 measured DLAs, is by
\cite{Storrie00}. At low redshift, DLAs are rare due to the low volume
of space surveyed by QSO sight-lines and the fact the spectral
observations must be made in the UV. To increase the number
statistics, \cite{Rao00} used a alternative approach by selecting
candidate DLAs from a sample of MgII absorption line systems. Although
$\lesssim$ 30 DLAs are now known at $z<1.65$, the number at $z<0.1$ is
still only two. Significant evolution of galaxies (and hence DLAs) is
still expected in the range $0<z<1.65$. Therefore DLAs are unlikely to
provide a suitable technique for measuring  $f(\nhi)$ at $z=0$.

Fortunately, at $z=0$ the alternative method of 21-cm emission can be
used to measure $f(\nhi)$. The 21-cm emission method has two main
advantages. Unlike QSO absorption observations, 21-cm emission does
not suffer from the low number statistics. Secondly, since the
properties of galaxies detected in 21-cm are known, they can be used to
infer the distribution of DLA galaxy properties such as luminosity,
surface brightness, and \HI\ mass. Hence we are able to assign the
probability that a particular galaxy type gives rise to a DLA at
$z=0$.

Previous calculations of the column density distribution function from
21-cm emission have been published by \cite{Rao93} and
\cite{Zwaan99,Zwaan02a}. The study by \cite{Rao93} used Arecibo
observations of 27 large spiral galaxies. \cite{Zwaan99,Zwaan02a} used
synthesis observations of optically-selected Ursa Major cluster
members and improved significantly on the previous calculation. The
benefit of using a galaxy cluster is that the spatial resolution is
the same for each galaxy. However, galaxy clusters are known to be
\HI-deficient \cite[e.g.][]{Solanes01} and may also have flatter \HI\
mass functions \cite[e.g.][]{Rosenberg02} which could bias the $f(\nhi)$
calculation.

To calculate $f(\nhi)$ from 21-cm emission we must be confident that
21-cm measurements of \nhi\ are equivalent to those from \Lya\
absorption.  According to \cite{Dickey90} ``The agreement between
ultraviolet and radio estimates of \nhi\ is excellent and somewhat
astonishing, considering that the angular resolution, experimental
techniques, and telescopes differ in all respects''.  This statement
refers to a comparison of high latitude Galactic 21-cm measurements
compared with \Lya\ absorption toward distant stars (in the same
direction) in the range $10^{20} \leq \nhi\ \leq 10^{21}$ \cm.  A
further issue is that high column density gas may clump on scales less
than the beam size. This concern has been addressed by \cite{Rao93} in
their calculation of the contribution of high \nhi\ gas to the \HI\
mass function (which is not affected by beam size) compared with the
integral of mass from the $f(\nhi)$ cross-section.  They find the
integral of mass in $f(\nhi)$ is equivalent to that from the mass
function at high \nhi. They concluded that any high \nhi\ column
density gas missed due to a large beam size is negligible.

In this paper we calculate $f(\nhi)$ from a randomly selected \HI\
sample of galaxies selected from the whole southern sky. This
selection process most closely mimics that of random QSO
sight-lines. The selection process and observations are described in
\S2. In \S3 $f(\nhi)$ at $z=0$ is calculated and compared to higher
redshift measurements. The data set is in then sorted by galaxy
properties and the contribution to the cross-section of \HI\ gas
satisfying the DLA criterion is calculated. In \S4 the implications of
the results are discussed and conclusions are presented in
\S5. Details of the galaxy sample, including \HI\ contour maps and
column density histograms are given in the appendix.


\section{Observations}

Galaxies for this study were selected from the \HI\ Parkes All-Sky
Survey (HIPASS) Bright Galaxy Catalogue (BGC). HIPASS is a blind
survey for \HI\ covering the entire southern sky. The data reduction
is described by Barnes et al. \citeyearpar{Barnes01}.  The 3$\sigma$
\HI\ mass limit of HIPASS is $\sim10^{6} D^{2}_{Mpc}\msun$, the 3$\sigma$
\HI\ column density limit is 4$\times10^{18}$ \cm\ and the FWHP
beamwidth is 15.5\arcmin. The BGC is the largest blind \HI\
extragalactic catalogue to-date. The galaxy finding and
parameterisation is described by Koribalski et
al. \citeyearpar{Koribalski02}. The \HI\ masses in the BGC were
calculated using Local Group corrected velocities and an $H_0$ = 75
\kms\ Mpc$^{-1}$.

HIPASS poorly resolves most galaxies and can usually only measure a
lower limit to the column density, equivalent to the \HI\ being
smoothly spread over the entire beam. To measure the distribution of
\HI\ column densities, higher resolution mapping is
required. Thirty-five galaxies were chosen at random from the BGC for
mapping with the Australia Telescope Compact Array (ATCA)\footnote{The
Australia Telescope Compact Array is part of the Australia Telescope
which is funded by the Commonwealth of Australia for operation as a
National Facility managed by CSIRO.}. To provide a representative
range of \HI\ mass, galaxies were chosen with an even mass
distribution in the range 7.5 $<$ log \mhi $<$ 10.5 \msun. Galaxies
were restricted to those with heliocentric velocities in the range 500
to 1700 \kms\ so that the spatial resolution remained comparable
across the sample.

Each galaxy was observed with the ATCA 375 and 750D array
configurations. The data set was reduced in MIRIAD, the velocity
resolution in each datacube is 3.3~\kms. The datacubes have an average
restored beam of 1\arcmin$\times$1\arcmin and typical RMS noise in
line-free regions of 3.7 mJy beam$^{-1}$, this corresponds to a
3$\sigma$ column density limit, over 3 velocity channels, of
2$\times10^{19}$~\cm.  The \HI\ column density (\nhi) of each pixel in
each galaxy was calculated by summing flux density along the spectral
axis. The integrated flux density, $Sdv$ (Jy \kms), was converted to a
brightness temperature in the usual manner, $T_b = 1.36\times10^3
\lambda^2 S/b_{maj} b_{min}$, where $\lambda$ is in cm and the major
and minor beam axes, $b_{maj}$ and $b_{min}$, in arcsec. $T_{b}$ is
converted to column density using $\nhi = 1.823\times 10^{18} \int
T_{b}dv$ cm$^{-2}$.  This calculation assumes that the gas is
optically thin.  \cite{Dickey90} used typical Galactic \HI\ parameters
(T${_s}\sim50$K and $\Delta v\sim10$\kms) to show that the optical
depth is greater than 1 for $\nhi > 10^{21}$ \cm. Consequently the
column density may be underestimated as \nhi\ approaches $10^{21}$
\cm, providing an upper limit to these observations. Regions of high
\nhi\ may also be underestimated due to the resolution of the
observations. Nearby galaxies which have been mapped at high
resolution show many regions with $\nhi> 10^{21}$ \cm \citep[e.g. the
LMC,][]{staveley03}. Table~A1 summarises the \HI\ parameters for each
of these galaxies. Also included are optical parameters from the
Lyon-Meudon Extragalactic Database (LEDA). Some galaxies are new or
recent discoveries and therefore some optical parameters are
unavailable. For example, HIPASS J0949-56 lies behind the Milky Way
and no optical counterpart is visible due to high foreground
extinction (A$_B$=9.2). Figure~\ref{fig:olays}1 gives the \nhi\
contours overlaid on a DSS image of each galaxy. An \nhi\ histogram
for each galaxy is given in Figure~\ref{fig:nhihist}2.


\section{Data Analysis}
\begin{figure}
\vspace{15pc}
\includegraphics{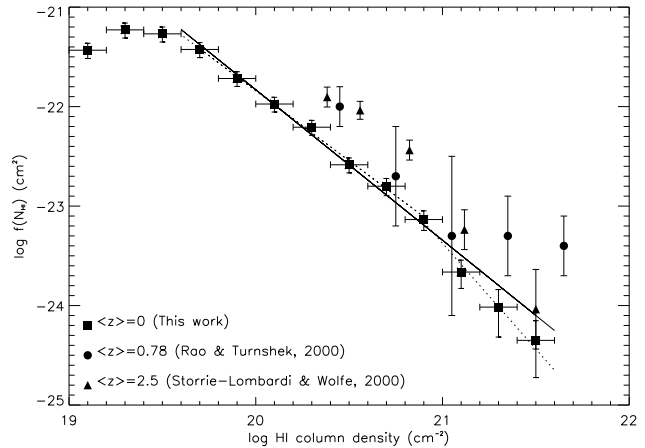}
\caption{Column density distribution function at $z=0$ for all
galaxies in the sample (data points). The error bars are
calculated using Poisson statistics. A comparison is made to the high
redshift ($\langle z\rangle= 2.5$) measurements of \citet{Storrie00}
(triangles) and intermediate redshift ($\langle z\rangle= 0.78$)
measurements of \citet{Rao00} (circles). The distribution is fitted with
a single power law (solid line) and double power law (dotted line).}
\label{fig:cddfall}
\end{figure}

The column density distribution function, from each galaxy ($i$) in
the sample, is calculated using
\begin{equation}
f(\nhi)=\frac{c}{H_0}\frac{1}{N}\sum^N_{i=1}\frac{\phi(\mhi)_iA(\nhi)_i}{d\nhi}
\hspace{8mm}[cm^{2}],
\end{equation}
where $A(\nhi)$ is the area in Mpc$^2$ occupied by column densities in
the range \nhi\ to $\nhi+d\nhi$ (d log\nhi = 0.2 dex is used
throughout). The function is normalised by the space density,
$\phi(\mhi)$ taken from the HIPASS BGC \HI\ mass function (Zwaan et
al. \citeyear{Zwaan02b}), which has a faint end slope of $\alpha =
-1.30\pm0.08$ and a characteristic \HI\ mass of
$\log(M^*/\msun)=9.79\pm0.06$ and normalisation
$\theta^*=(8.6\pm2.1)\times10^{-3}$ Mpc$^{-3}$. The resulting
distribution function is plotted in Figure~\ref{fig:cddfall}. A power
law, $f(\nhi) = B\nhi^{-\beta}$, is fitted with $\beta=1.5\pm0.1$ in
the range $19.6<\log \nhi<21.6$. The three highest column density
points do not agree well with this power law. A double power law is
therefore fitted with a break at $\log \nhi=20.9$. In this case the
lower column densities, $19.6<\log \nhi<20.9$, are best fitted with
$\beta=1.4\pm0.2$ and the upper column densities, $20.9<\log
\nhi<21.6$, with $\beta=2.1\pm0.9$. The double power law is also given
in Figure~\ref{fig:cddfall}.

The $z=0$ $f(\nhi)$ from 21-cm is compared to higher redshift
measurements from QSO absorption line studies. The high redshift
($\langle z\rangle\sim 2.5$) DLA measurements are taken from
\cite{Storrie00} and the intermediate redshift ($\langle z\rangle\sim
0.78$) DLA measurements from \cite{Rao00}. (Although $q_0$ is
unimportant at $z=0$, the high redshift $f(\nhi)$ and \dndz\ use
$q_0=0$.) We find that with decreasing redshift, $f(\nhi)$ decreases
at all column densities and flattens for $20.9<\log \nhi<21.6$ from
$\langle z\rangle\sim 2.5$ to 0. Quantitatively, the flattening is
determined by fitting a power law in this high \nhi\ range to the
$\langle z\rangle\sim 2.5$ ($\beta=2.5$) and $z=0$ ($\beta=2.1$)
distribution functions. The flattening of $f(\nhi)$ at high column
densities is attributed to growth of denser objects (i.e. galaxies),
therefore reducing the number of lower column density systems and
increasing the high \nhi\ systems. The overall decrease in $f(\nhi)$
to $z=0$ can be explained by the consumption of \HI\ gas to form
stars.  The magnitude of the \cite{Rao00} $f(\nhi)$ fits with the
general evolutionary picture. However the slope of the function is not
in good agreement, although the error bars are large enough for the
above description to hold within the stated uncertainties.

The departure from the power law at $\nhi\ < 10^{20}$\cm\ is perhaps
partly due to the \HI-\HII\ transition, described as the ``footprint''
of ionisation by \cite{Corbelli01}, essentially marking the \HI\ edge
of the galaxies. It may also be due to the detection limit of our
observations. The detection limit is a function of the line width, so
measurements of low column density regions of galaxies with large
line widths are most likely underestimated.


\begin{figure}
\vspace{15pc}
\includegraphics{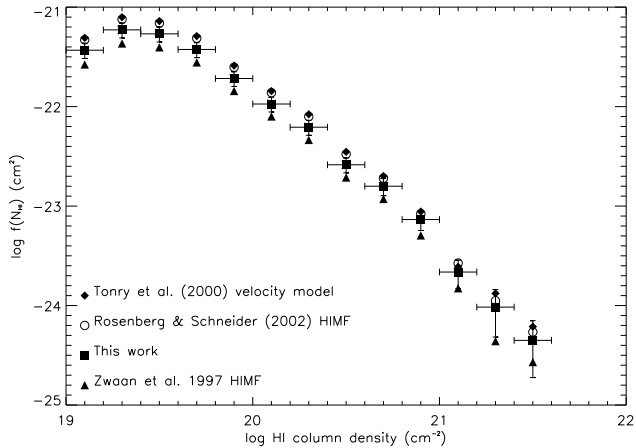}
\caption{The effect of systematic errors on the column density
distribution function. The original function is included along with
two different \HI\ mass functions \citep{Zwaan97,Rosenberg02} and a
velocity flow model to calculate the distance to each galaxy
\citep{Tonry00}. The different functions show that systematic
uncertainties are small and have little effect on the shape of
$f(\nhi)$.}
\label{fig:cddfcomp}
\end{figure}

Systematic uncertainties in $f(\nhi)$ could arise from errors in the
galaxy distances or the \HI\ mass function. The distance to each
galaxy was calculated using the radial velocity, corrected for the
motion of the Local Group. Alternatively, a parametric model for the
velocity field of galaxies can be used to calculate distance. The
model used is by \cite{Tonry00}, uncertainties in the distance affects
the $\phi(\mhi)$ and $A(\nhi)$ terms in $f(\nhi)$. The difference
between the two distance calculations is $\sim20\%$ for most galaxies
in the sample. The resulting $f(\nhi)$ is plotted in
Figure~\ref{fig:cddfcomp} along with the original calculation. In
addition, $f(\nhi)$ is calculated with two different \HI\ Mass
functions, one with shallower slope \citep{Zwaan97} and one with a
stepper slope \citep{Rosenberg02}. These systematic uncertainties
have little effect on the shape and normalisation of the column
density distribution function beyond the random errors already quoted.


\begin{figure}
\vspace{15pc}
\includegraphics{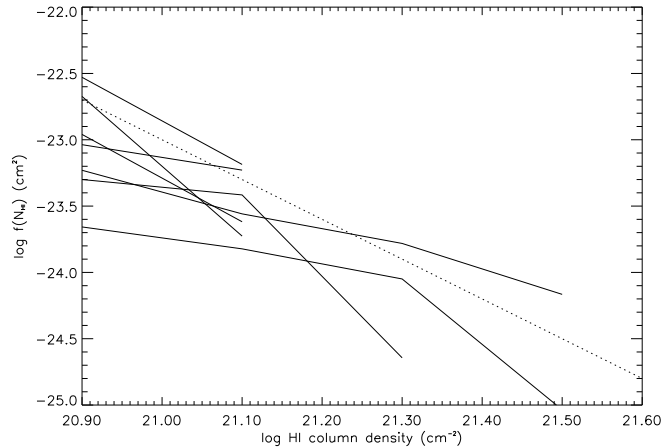}
\caption{Column density distribution function for individual spiral
galaxies only (solid lines). The dotted line, \nhi$^{-3}$, is
predicted theoretically (arbitrary normalisation).}
\label{fig:slope}
\end{figure}

The power law slope of the column density distribution is
theoretically predicted to be $-3$ at the high \nhi\ end for disc
galaxies \citep{Milgrom88, Wolfe95}. This prediction together with
a measured $f(\nhi)$ slope flatter than $-3$ has been used to argue
that disc galaxies cannot be responsible for all DLA systems
\citep{Rao00}. Considering only the spiral galaxies in the sample, we
find that f(\nhi), for \nhi $\geq 10^{21}$ \cm, has a power law slope
of $-2.6\pm1.0$. In Figure~\ref{fig:slope}, f(\nhi) for each spiral
galaxy contributing to this \nhi\ range is shown. If we consider both
spiral and non-spiral galaxy types, the power law slope in this high
\nhi\ region increases to $-2.1\pm0.9$. Thus, including non-spiral
galaxies in the sample, the data still agrees, within the
uncertainties, with the theoretical prediction for spiral galaxies
only.

\begin{figure}
\vspace{15pc}
\includegraphics{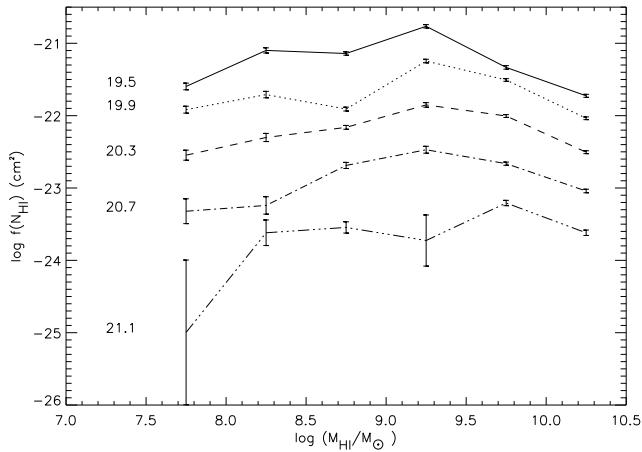}
\caption{Column density distribution function at $z=0$
as a function of galaxy \HI\ mass, plotted at six different column densities.}
\label{fig:himass}
\end{figure}

To emphasise the relationship between $f(\nhi)$ and total galaxy \HI\
mass these are plotted in a range of column density bins in
Figure~\ref{fig:himass}. $f(\nhi,\mhi)$ has the same basic shape at
all column densities. For most \nhi, $f(\nhi,\mhi)$ is essentially a
flat function of \HI\ mass, rising slightly from log \mhi\ $=9$ to 9.5
\msun\ and dropping off sharply after log \mhi\ $=10$ \msun. To
compare with DLA absorption line statistics, we also integrate
$f(\nhi)$ from \nhi~$\geq2\times10^{20}$\cm, which is equivalent to
calculating the cross-section of \HI\ satisfying the DLA criteria, or
the number density of DLAs per unit redshift.

\begin{equation}
\frac{dN_{DLA}}{dz}=\int_{\nhi=2\times10^{20}}^{\nhi_{max}}f(\nhi)d\nhi
\end{equation}

\noindent For the whole sample, \dndz\ =
0.058$\pm0.006$ (random) $^{+0.03}_{-0.02}$ (systematic) at $z=0$. The random
uncertainty is based on Poisson statistics and the systematic
uncertainty is from the \HI\ mass function. This value agrees very
well with the parameterisation, \dndz$=0.055(1+z)^{1.11}$ over a large
redshift range by \cite{Storrie00} (see their Figure 11). This results
confirms the suggestion by \cite{Storrie00} that there is no intrinsic
evolution in the product of space density and cross-section of DLAs
with redshift to $z=0$.

\begin{figure}
\vspace{15pc}
\includegraphics{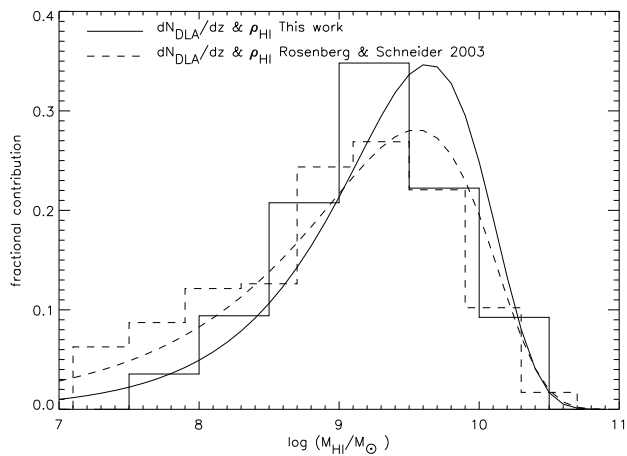}
\caption{Fractional contribution of each \HI\ mass bin to \dndz\
  (histogram) and \HI\ mass density, $\rhohi$ (\msun $Mpc^{-3}$,
  continuous line) comparing the work in this paper (solid histogram
  and line) with Rosenberg \& Schneider (2003) (dashed histogram and
  line).}
\label{fig:hidensity}
\end{figure}

\begin{figure}
\vspace{15pc}
\includegraphics{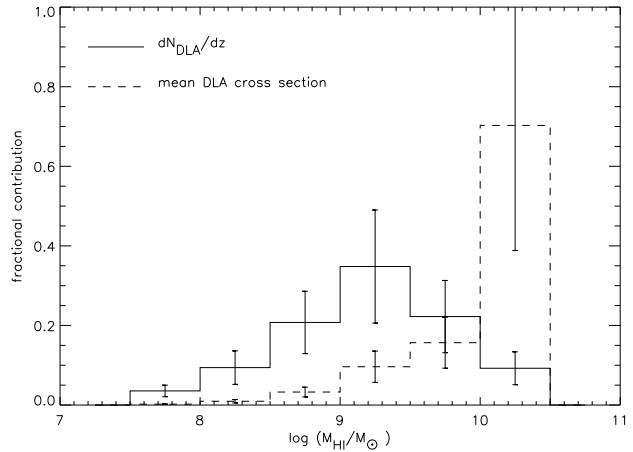}
\caption{Fractional contribution of each \HI\ mass bin to \dndz\
  (solid line) and the mean DLA cross-section (dashed line).}
\label{fig:mass}
\end{figure}

Figure~\ref{fig:hidensity} compares the fractional contribution of
\dndz\ to the \HI\ mass density, \rhohi=$\phi(\mhi)\times\mhi$ as a
function of log \mhi. The histogram shows that 34\% of \dndz\ can be
attributed to galaxies with log \mhi\ $<9.0$. This is in contrast to
\rhohi\ (continuous line in figure), which is dominated by $M^*$
galaxies. In comparison 26\% of \rhohi\ is due to galaxies with log
\mhi $< 9.0$. Furthermore, Figure~\ref{fig:mass} shows that the mean
DLA cross-section of log \mhi\ $<9.0$ galaxies is a mere 4\%.

Although the cross-section of \HI\ satisfying the DLA criterion in an
individual low \HI\ mass galaxy may be small, the product of cross
section and space density is more significant than the product of \HI\
mass and space density. This leads to \dndz\ as a function of \HI\
mass peaking at less than M$_*$. Low mass galaxies may contribute
little to the \HI\ mass density, however they make up an important part
of \dndz.

This result is similar to \cite{Rosenberg03} if the different \HI\
mass functions are considered. In both cases the \dndz\ histogram
generally sits slightly above \rhohi\ for $\mhi < M^*$ and slightly
below for $\mhi > M^*$. \cite{Rosenberg03} find a tight correlation
between log \mhi\ and log $A(\nhi \geq 2\times10^{20} cm^{-2})$ and
use this to calculate \dndz. In that case \rhohi\ $\propto$ \dndz, so
the shape of \rhohi(\mhi) should match \dndz(\mhi) peaking at
$M^*$. In practice the shape of \dndz\ deviates slightly from \rhohi\
since \cite{Rosenberg03} use the $1/V_{tot}$ value for each individual
galaxy rather than $\phi$(\mhi) from the function form of the \HI\
mass function.


\subsection{Relationship between DLA-gas and Optical Properties}

\begin{figure}
\vspace{15pc}
\includegraphics{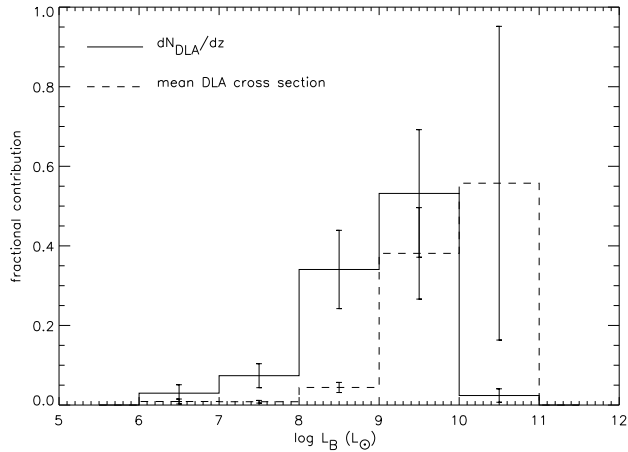}
\caption{Fractional contribution of each luminosity decade to \dndz\
  (solid line) and the mean DLA cross-section (dashed line).}
\label{fig:lum}
\end{figure}

\begin{figure}
\vspace{15pc}
\includegraphics{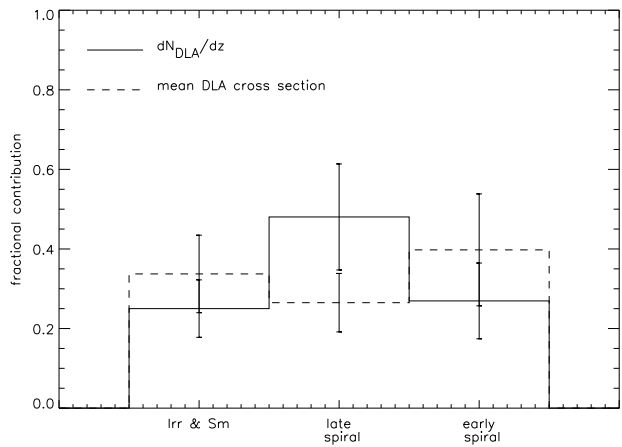}
\caption{Fractional contribution of each morphological type to \dndz\
  (solid line) and the mean DLA cross-section (dashed line).}
\label{fig:morph}
\end{figure}

\begin{figure}
\vspace{15pc}
\includegraphics{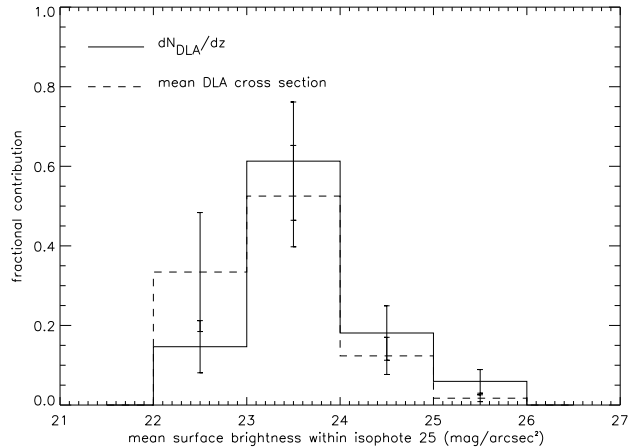}
\caption{Fractional contribution of each mean surface brightness
  decade to \dndz\ (solid line) and the mean DLA cross-section (dashed
  line).}
\label{fig:sb}
\end{figure}

\begin{figure}
\vspace{15pc}
\includegraphics{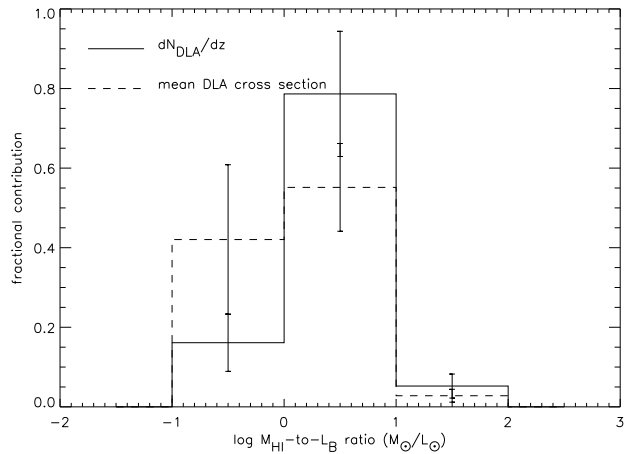}
\caption{Fractional contribution of each mass-to-light ratio decade to
  \dndz\ (solid line) and the mean DLA cross-section (dashed line).}
\label{fig:mlratio}
\end{figure}

In this section we explore the relationship between \dndz\ and optical
properties of galaxies in the sample. \dndz\ is still normalised by
the \HI\ mass function, but is weighted to reflect the number
distribution of optical properties in the sample. The mean DLA
cross-section, $A(\nhi \geq 2\times10^{20} cm^{-2})$, for galaxies
with each different property is also calculated. The fractional
contributions to \dndz\ and mean DLA cross-section as a function of
each galaxy property is given in Figures~\ref{fig:lum} to
\ref{fig:mlratio}. In each case the \dndz\ contribution (solid
histogram) describes the probability of a galaxy with a particular
property giving rise to a DLA system. The mean DLA cross-section
(dashed histogram) indicates for a galaxy with a particular property,
its non-weighted DLA cross-section on the sky.

First, the likelihood of a galaxy in different luminosity ranges
giving rise to a DLA system is calculated. Galaxies in the range 8
\lsun\ $<$ log $L_B <$ 10 \lsun\ dominate \dndz.  Sub-$L_*$ (log $L_B
<$ 9 \lsun) galaxies account for 45\% of the DLA cross-section. In
contrast these sub-$L_*$ galaxies have very small (6\%) DLA
cross-sections compared with $L>L_*$ galaxies. This has interesting
consequences for the interpretation of DLA-galaxy associations. For
example, although sub-$L_*$ galaxies are just as likely to give rise
to a DLA absorber, given a DLA system in a field with one sub-$L_*$
and one super-$L_*$ galaxy, the super-$L_*$ galaxy is $\sim$10 times
more likely to be responsible for the DLA by virtue of its much larger
\HI\ cross-section.

To investigate the relationship between \dndz\ and galaxy morphology,
the sample was divided into three parts. The first group contains
Irregular and Magellanic-type galaxies. The second group, late type
spirals, includes Sd and Sc types. The early type spiral group is
comprised of Sb, Sa and S0 types. We find that \dndz\ is dominated by
late type spirals, which contribute 48\% to the cross-section of
DLA-gas on the sky. The contribution of Irregular and Magellanic types
is non-negligible at 25\%. Individually however, early type spirals
each have the largest DLA cross-section (40\%).

\dndz\ as a function of galaxy mean surface brightness within the
magnitude 25 isophote (\surf) is dominated by galaxies with 23 $<$
\surf $<$ 24 mag arcsec$^{-2}$. The mean DLA cross-section of these
galaxies also dominate the surface brightness division. Lower surface
brightness galaxies, with \surf $>$ 24 mag arcsec$^{-2}$, make up 22\%
of the cross-section. Five galaxies did not have this parameter
available in the LEDA database. One of these is HIPASS J0949-56, which
lies behind the Milky Way. Another, HIPASS J1320-21 is a high surface
brightness S0 galaxy. The other three (HIPASS J0905-56, J1321-31 and
J2222-48) all appear to be LSB galaxies (see Figure~A1). The fact that
these galaxies are missing from the analysis introduces a bias against
low surface brightness galaxies. Including these three galaxies in the
lowest surface brightness bin (25 $<$ \surf\ $<$ 26 mag arcsec$^{-2}$)
increases the \dndz\ contribution from 3 to $6\pm5\%$.

The relationship between \dndz, \HI\ mass and $L_B$ is
combined in the calculation of \mhi-to-L$_{B}$ ratio. The cross
section of DLA-gas is dominated by gas-rich galaxies with
\mhi-to-L$_{B}$ ratios in the range 1 -- 10. The very high
\mhi-to-L$_{B}$ ratio galaxies contribute little to the DLA cross
section (5\%), mostly due to their rarity. It is also
due to the fact that most of the \HI\ in these galaxies has a low column
density. The contribution of very high \mhi-to-L$_{B}$ ratio galaxies
to $dN/dz$ for log \nhi\ $\leq 19.1$ \cm\ is $18\pm5\%$.


\section{Discussion}

The flattening of f(\nhi) between $\langle z\rangle= 2.5$ and $z=0$
for $\nhi\gtrsim10^{20.5}$\cm\ can be explained by accreting clumps of
\HI\ into the central regions of galaxies and DLA systems. Milgrom
(1988) argues that the slope of the column density distribution
function reflects the density distribution within the absorbers
themselves. A case is presented where lines of sight through spherical
gas clouds give rise to f(\nhi)$\propto\nhi^{-\beta}$.  The column
density profile of each sphere is given by $\nhi(l) =
N_o[1+(l/r_c)^2]^{-c}$, where $l$ is the impact parameter and $r_c$ is
the core radius. The power law indices of the local and global
distribution are related through $\beta=(c+1)/c$. The high redshift
column density distribution \citep{Storrie00}, for $\nhi\ge
10^{20.5}$\cm, has a power law index of $\beta = 2.2$. Comparing
$\beta=2.2$ (with $c=0.8$) to our $z=0$ measurement of $\beta=1.6$
(with $c=1.7$), yields a more centrally concentrated column density
profile for galaxies (or DLAs) at $z=0$. This picture lends support to
the hierarchical construction of galaxies and DLAs. The alternative
model, monolithic collapse into disc-like systems, predicts $f(\nhi)
\propto \nhi^{-3}$ for very high column densities (\nhi $\geq 10^{21}$
\cm).  Our data is marginally consistent with this theoretical
prediction for disc galaxies. However, we are unable to confirm a
shallower power law for non-disc galaxies, because of a lack of
measured high column densities, due to resolution limitations.

The number density of DLA systems per unit redshift is calculated to
be \dndz\ = 0.058$\pm0.006$ (random) $^{+0.03}_{-0.02}$ (systematic)
at $z=0$. This result agrees well with an extrapolation of higher
redshift evolution measured by \cite{Storrie00}. Our result is in
moderate agreement (within the large uncertainties) with
\cite{Churchill01} who finds \dndz\ = $0.08^{+0.09}_{-0.05}$ at
$\langle z\rangle= 0.06$ from MgII QSO absorption systems. Our result
is also consistent with the 21-cm studies by \cite{Zwaan02a} and
\cite{Rosenberg03} who find \dndz\ = 0.042$\pm$0.015 and
0.053$\pm$0.013 respectively.

The association of \Lya\ absorption lines with particular galaxies is
a subject of much debate. Some statistical studies compare galaxy
survey data complete to a limit of L$_*$ with \Lya\ absorption systems
\citep[e.g.][]{Penton02}. We have shown statistically, that sub-L$_*$
galaxies account for 45\% of the DLA cross-section at $z=0$. If such a
large proportion of galaxies are missed, the strength of the
galaxy-absorber cross-correlation function may be underestimated.

The results presented here agree with the targeted searches for low
redshift DLA galaxies that find a range of morphologies and
luminosities. Unfortunately 21-cm emission measurements of galaxies
are limited to $z<0.1$, whereas the detection \Lya\ absorption is
bounded only by the brightness of the background probe. However, we
can project these results to redshifts beyond zero. We have shown that
low \HI\ mass, low luminosity galaxies contribute significantly to the
DLA cross-section at $z=0$. These types of galaxies are more common at
higher redshifts \citep[e.g.][]{Ellis97}. Provided the DLA
cross-section of these types of galaxies do not get significantly
smaller with redshift, it is likely they also account for an important
part of the high redshift DLA cross-section.

In future studies, where the absorption-line statistics and properties
of $z=0$ DLA galaxies are better known, f(\nhi) and \dndz\ from 21-cm
emission will provide a test of possible biases in the DLA galaxy
population. Selection effects include gravitational lensing of
QSO-galaxy alignments, bringing otherwise dimmer QSOs into a
statistical sample and creating a `by-pass' effect where the
line-of-sight avoids the galaxy centre \citep{Smette97}. Although,
\cite{Lebrun00} showed that lensing is not important in the sample of
DLA galaxies presented in \cite{LeBrun97}, they note that brighter
QSOs may be affected. Spiral galaxies could be under-represented in
the DLA galaxy population due to their large dust content, obscuring
any background QSO \citep{Ostriker84}. A recent study of radio loud
QSOs \citep{Ellison01} found that a dust-induced bias in optical QSO
surveys may have led to an underestimate of $\Omega_{DLA}$ and \dndz\
by at most a factor of two. However, DLAs are 4 times as likely to be
found in the proximity of radio loud QSOs \citep{Ellison02} which
could overestimate $\Omega_{DLA}$ and \dndz. If the DLA cross-section
from both 21-cm and QSO absorption is well measured and compared these
biases can be better understood.


\section{Conclusions}
We have calculated the column density distribution function at $z=0$
from an \HI-selected sample of galaxies. We find that $f(\nhi)$
follows a power law with slope = $-1.5\pm0.1$ in the range $19.6<\log
\nhi<21.6$. When compared to high redshift QSO absorption line data,
$f(\nhi)$ flattens and decreases with evolution to $z=0$. This
evolution can be interpreted as accretion of lower column density
systems onto higher ones, thus reducing the number of low column
density galaxies. This picture fits with the hierarchical formation
model of galaxies and DLA systems. Conversely, we are in marginal
agreement with the theoretical prediction $f(\nhi) \propto\nhi^{-3}$
for very high column densities in spiral discs. The fact that we find
$f(\nhi)\propto\nhi^{-2.1\pm0.9}$ for very high column densities, for
all types of galaxies, including spiral discs means that we cannot use
this analysis to rule out the alternative model, that spiral discs are
responsible for all DLA systems.

However, analyzing the the slope of the $f(\nhi)$ power law is not the
only method of determining the likely galaxy types which gives rise to
DLA systems. The calculation of \dndz\ as function of \HI\ mass,
luminosity, galaxy morphology, \mhilb\ ratio and surface brightness
reveals a range of galaxy types contribute to the cross-section of
\nhi\ satisfying the DLA criteria. In particular we find that galaxies
with $\log \mhi< 9.0$ make up 34\% of \dndz; Irregular and Magellanic
types contribute 25\%; galaxies with surface brightness, \surf $>$ 24
mag arcsec$^{-2}$ account for 22\% and sub-$L_*$ galaxies contribute
45\% to \dndz. These results agree with findings from the imaging of
low redshift DLA galaxies. Integrating over all galaxies in the sample
we find that \dndz$=0.058$, which is consistent with no intrinsic
evolution in the product of space density and cross-section of DLAs
with redshift. This result agrees with the redshift evolution
predicted from high and intermediate DLA systems \citep{Storrie00} and
is moderately consistent with very low redshift QSO absorption line
studies \citep{Churchill01}.

\bibliographystyle{mn2e}
\bibliography{mn-jour,cddfref}

\appendix

\section[]{Galaxy Sample}

Includes a table of all the galaxies, an optical image with \nhi\
contours overlaid and a column density histogram for each galaxy.

\begin{figure*}
\vbox to 470mm{
\includegraphics{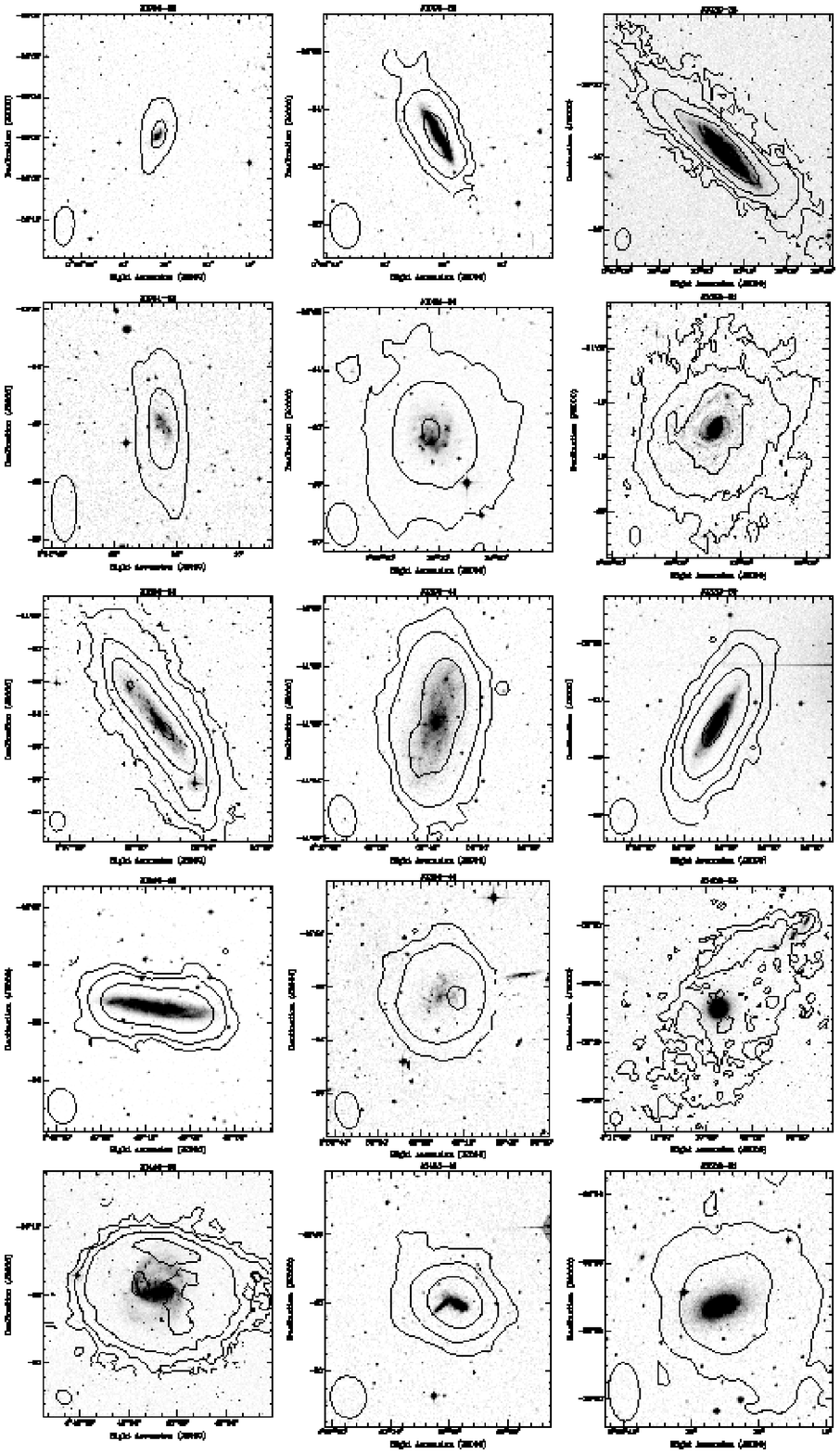} 
{\vfil {\bf{Figure A1.}} Optical images of galaxies in the sample with
  \HI\ column density contours overlaid at log \nhi\ = 19.9, 20.3,
  20.7, 21.1 \& 21.5 \cm. The restored beam is given in the bottom
  left-hand corner of each image.}
\vfil}
\label{fig:olays}
\end{figure*}

\begin{figure*}
\vbox to 470mm{
\includegraphics{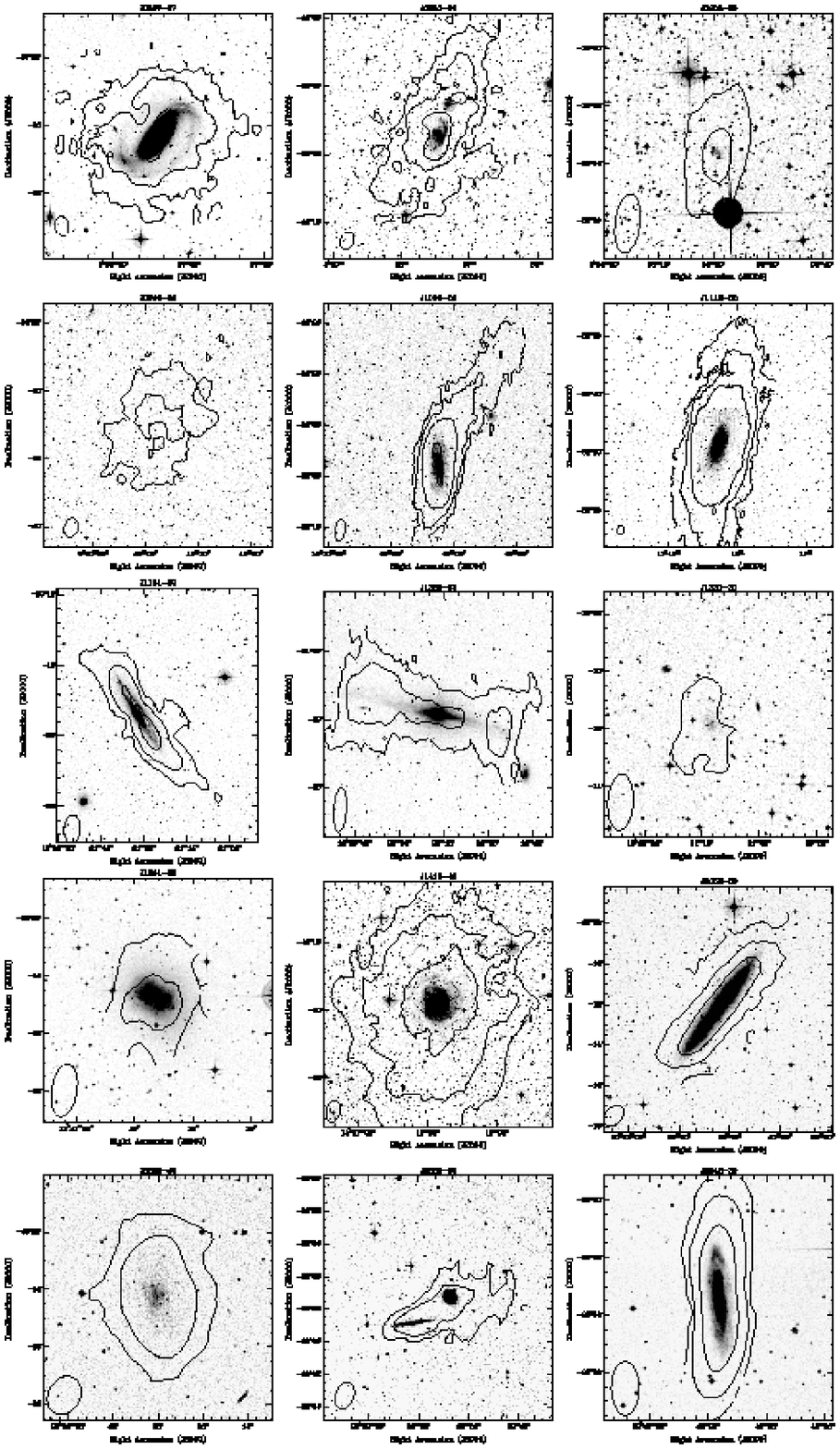} 
{\vfil {\bf{Figure A1 cont.}} Optical images of galaxies in the sample with
  \HI\ column density contours overlaid at log \nhi\ = 19.9, 20.3,
  20.7, 21.1 \& 21.5 \cm. The restored beam is given in the bottom
  left-hand corner of each image.}
\vfil}
\end{figure*}

\begin{figure*}
\vbox to 190mm{
\includegraphics{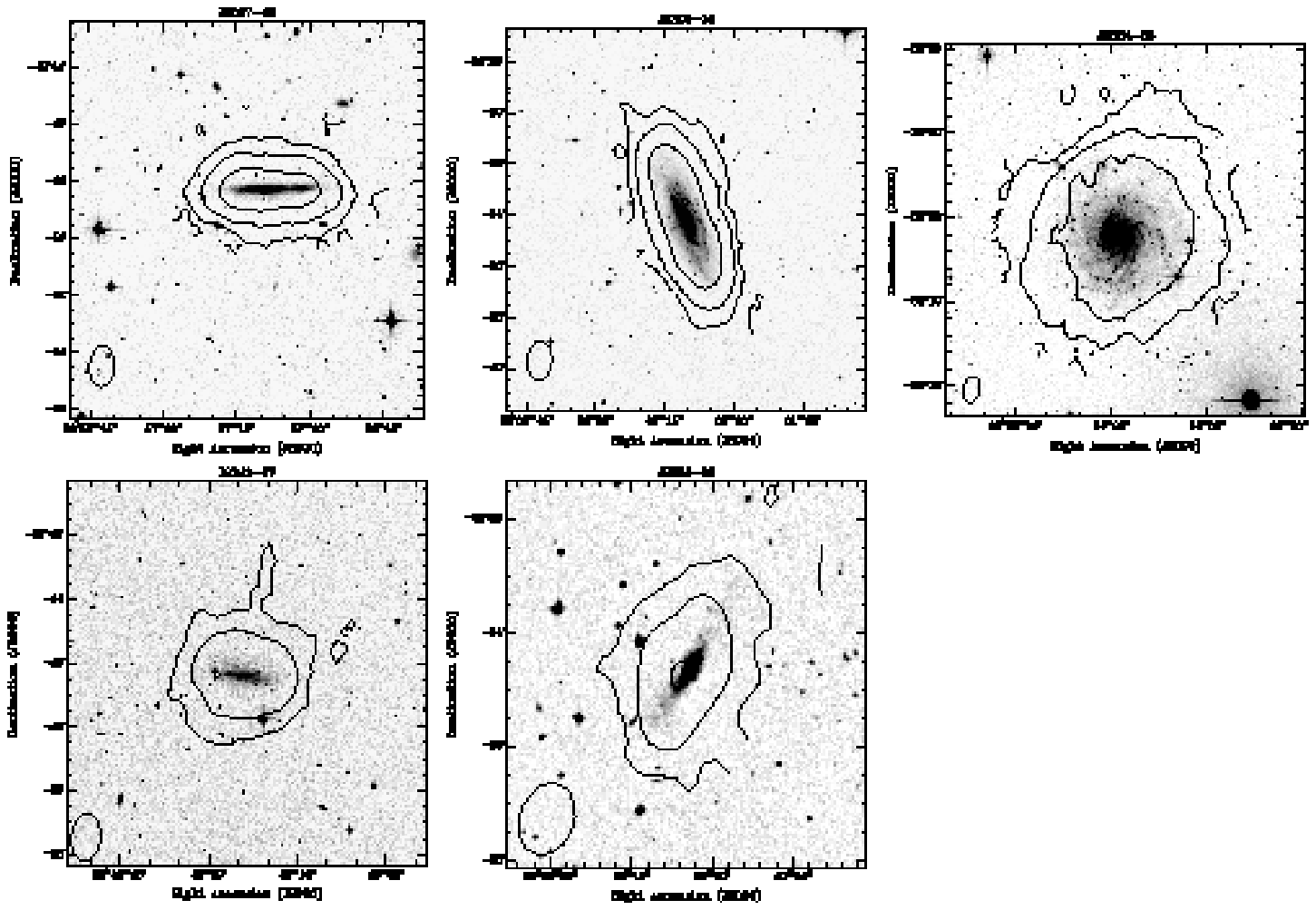} 
{\vfil {\bf{Figure A1 cont.}} Optical images of galaxies in the sample with
  \HI\ column density contours overlaid at log \nhi\ = 19.9, 20.3,
  20.7, 21.1 \& 21.5 \cm. The restored beam is given in the bottom
  left-hand corner of each image.}
\vfil}
\end{figure*}

\begin{table*}
\centering
\begin{minipage}{160mm}
\begin{tabular}{llrrrrrrl}
HIPASS    &  Optical ID      & V$_{hel}$ & Distance & log \mhi\  & beam &
\surf &  \mhi/$L_{B}$  &  Morphology  \\
Name& & (\kms) &(Mpc) & (\msun) &($\arcsec\times\arcsec$(kpc))&  (mag \arcsec$^{-2}$)&  (\msun/\lsun)&\\
  
\hline
J0005-28 &  ESO409-IG015    &  736  &   10.1 &  8.25  & 118$\times$62(3.1)&24.0  &   2.2   &   Sc   \\                                
J0008-29 &  NGC0007         & 1508  &   20.4 &  9.15  &  96$\times$67(6.6)&  23.7  &   2.2   &   SBc  \\			       
J0030-33 &  NGC0134         & 1582  &   20.9 & 10.16  &  91$\times$66(6.8)&  23.5  &   1.3   &   SBbc \\			       
J0031-22 &  ESO473-G024     &  540  &    7.6 &  7.99  & 140$\times$57(2.1)&  24.1  &   5.4   &   Irr  \\			       
J0034-30 &  UGCA006         & 1582  &   21.0 &  9.25  &  92$\times$66(6.8)&  24.0  &   2.9   &   SBm  \\			       
J0052-31 &  NGC0289         & 1629  &   21.5 & 10.24  & 106$\times$65(6.8)&  23.9  &   2.5   &   SBbc \\ 			       
J0256-54 &  ESO154-G023     &  574  &    5.3 &  8.97  &  70$\times$60(1.6)&  24.0  &   6.8   &   SBm  \\			       
J0309-41 &  ESO300-G014     &  955  &   10.8 &  8.86  &  90$\times$58(3.1)&  24.4  &   2.0   &   SBm  \\			       
J0333-50 &  IC1959          &  640  &    6.1 &  8.37  &  74$\times$60(1.8)&  23.1  &   1.7   &   SBm  \\			       
J0349-48 &  IC2000          &  981  &   10.5 &  8.92  &  75$\times$62(3.2)&  23.2  &   1.2   &   SBc  \\			       
J0359-45 &  Horologium dwarf & 898  &    9.5 &  8.60  &  86$\times$57(2.6)&  ...   &   4.9   &   IB(s)\\			       
J0409-56 &  NGC1533         &  785  &    7.6 &  8.97  &  68$\times$65(2.4)&  22.8  &   0.9   &   S0   \\			       
J0445-59 &  NGC1672         & 1331  &   14.6 & 10.00  &  75$\times$59(4.3)&  22.9  &   0.8   &   SBb  \\			       
J0457-42 &  ESO252-IG001    &  657  &    5.9 &  7.95  &  74$\times$67(1.9)&  23.7  &   3.1   &   merger\\			       
J0506-31 &  NGC1800         &  809  &    8.1 &  8.28  & 103$\times$57(2.3)&  22.9  &   0.6   &   SBd  \\			       
J0507-37 &  NGC1808         &  995  &   10.4 &  9.25  &  86$\times$66(3.4)&  23.3  &   0.4   &   SBa  \\			       
J0857-69 &  ESO060-G019     & 1443  &   15.5 &  9.47  &  78$\times$62(4.7)&  23.3  &   1.9   &   SBcd \\			       
J0905-36 &  new             &  888  &    8.0 &  7.93  & 124$\times$56(2.2)&  ...   &  ...    &   Im   \\			       
J0949-56 &  HIZSS059\footnote{Henning et al. \citeyearpar{Henning00}}& 1762 & 19.6 &  9.58  & 85$\times$66(6.3)& ...&...&...\\      
J1009-29 &  NGC3137         & 1104  &   11.0&   9.56  & 134$\times$59(3.2)&  23.9  &   2.6   &   SBc  \\			       
J1118-32 &  NGC3621         &  730  &    6.2&   9.91  &  95$\times$79(2.4)&  23.5  &   1.9   &   SBcd \\			       
J1131-30 &  NGC3717         & 1730  &   19.7&   9.73  & 117$\times$75(7.2)&  23.9  &   1.4   &   Sb   \\			       
J1320-21 &  NGC5084         & 1721  &   20.7&  10.01  & 197$\times$59(6.0)&  ...  &   1.2   &   S0   \\			        
J1321-31 &  new\footnote{Banks et al. \citeyearpar{Banks99}}&  571 &5.0 & 7.54  &119$\times$ 57(1.4)& ...&11.2&Sm\\		       
J1341-29 &  NGC5264         &  478  &    4.0&  7.68  &  112$\times$56(1.1)& 23.1  &   0.3   &   Irr  \\			        
J1410-43 &  NGC5483         & 1771  &   21.1& 10.09  &  102$\times$63(6.4)& 23.6  &   2.0   &   SBc  \\			        
J2052-69 &  IC5052          &  584  &    5.9&  8.93  &   71$\times$50(1.4)& 22.6  &   1.3   &   SBcd \\			        
J2222-48 &  ESO238-G005     & 706  &     8.9& 8.47  &    85$\times$70(3.0)&...   &  50.2   &   Irr  \\			        
J2223-28 &  NGC7259         & 1780  &   24.5&  9.56  &  104$\times$70(8.4)& 22.6  &   4.6   &   Sb   \\			        
J2242-30 &  NGC7361         & 1249  &   17.3&  9.47  &  113$\times$62(5.2)& 23.4  &   1.6   &   Sc   \\			        
J2257-42 &  NGC7412A        &  930  &   12.1&  8.79  &   86$\times$59(3.5)& 24.5  &   4.6   &   SBd  \\			        
J2302-39 &  NGC7456         & 1199  &   15.9&  9.42  &   91$\times$62(4.8)& 23.9  &   1.3   &   Sc   \\			        
J2334-36 &  IC5332          &  701  &    9.3&  9.51  &   97$\times$60(2.7)& 24.6  &   1.4   &   SBcd \\			        
J2349-37 &  ESO348-G009     &  648  &    8.4&  8.35  &   90$\times$59(2.5)& 25.2  &  18.1   &   Irr  \\			        
J2352-52 &  ESO149-G003     &  576  &    6.5&  7.84  &   79$\times$60(1.9)& 23.9  &   1.8   &   Irr  \\			        
\hline
\end{tabular}
\caption{\HI\ parameters for each galaxy from HIPASS and optical
  parameters from LEDA. The columns are the HIPASS name, optical ID,
  heliocentric velocity, distance (calculated using Local Group
  corrected velocities), \HI\ mass, beam size of 21-cm map with the
  minor axis in kpc in parentheses, mean surface brightness within the
  25th magnitude isophote, \mhilb\ ratio and galaxy morphology.}
\end{minipage}
\end{table*}

\begin{figure*}
\vbox to 400mm{
\includegraphics{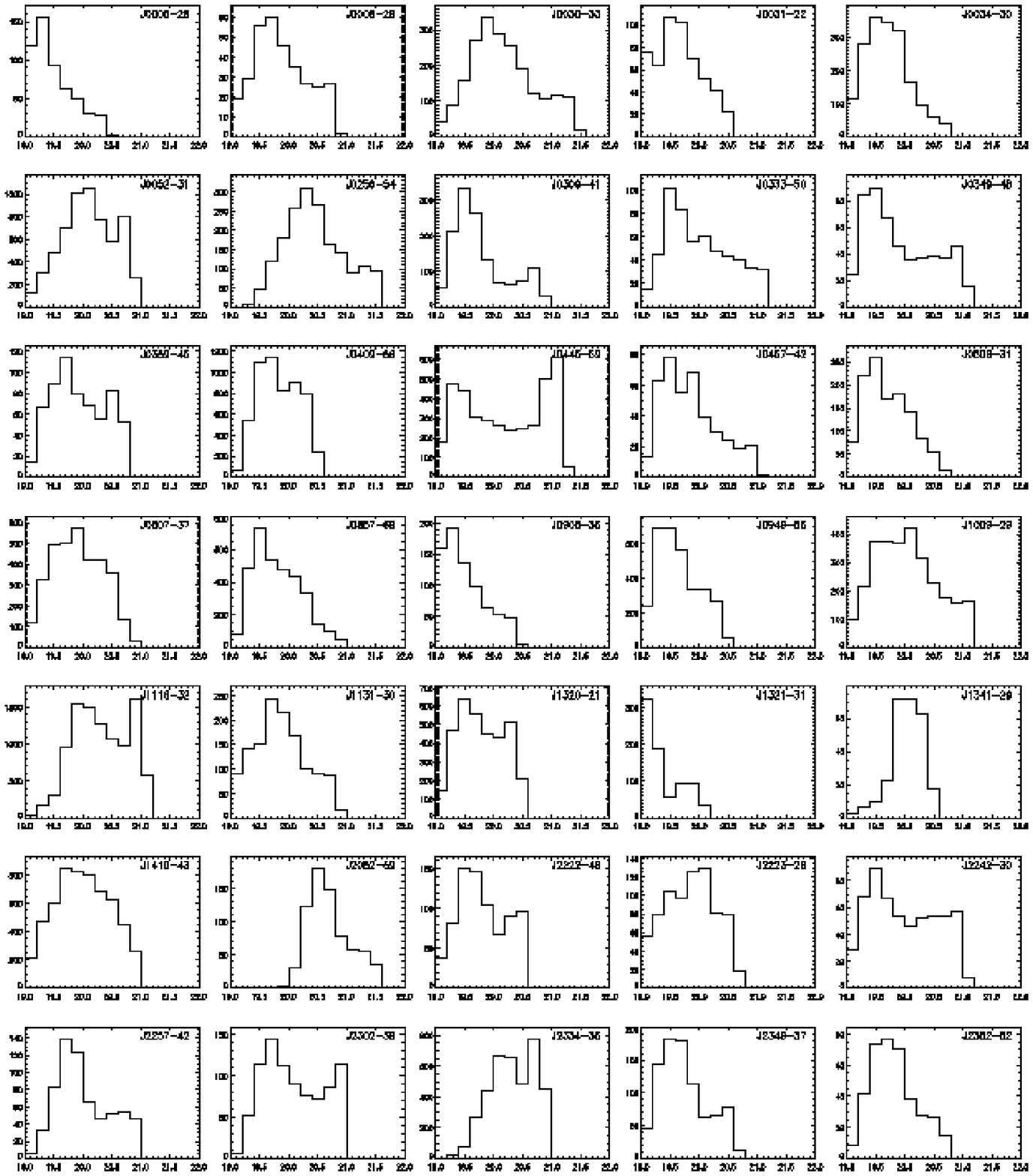} 
{\vfil {\bf{Figure A2.}} \nhi\ histograms for all galaxies in the
  sample. In each case, the x axis is log \nhi\
  (cm$^{-2}$) and the y axis is the number of pixels.}
\vfil}
\label{fig:nhihist}
\end{figure*}

\bsp

\label{lastpage}

\end{document}